# A Secure Dynamic Job Scheduling on Smart Grid using RSA Algorithm

P.Radha Krishna Reddy1, Ashim Roy     G.Sireesha, Ismatha Begum, S.Siva Ramaiah

*M.Tech Student*     *Assistant Professor*

*JNTU,Anantapur*     *JNTU,Hyderabad*

1pallavali@gmail.com

*Abstract*— Grid computing is a computation methodology using group of clusters connected over high-speed networks that involves coordinating and sharing computational power, data storage and network resources. Integrating a set of clusters of workstations into one large computing environment can improve the availability of computing power. The goal of scheduling is to achieve highest possible system throughput and to match the application need with the available computing resources. A secure scheduling model is presented, that performs job grouping activity at runtime. In a Grid environment, security is necessary because grid is a dynamic environment and participates are independent bodies with different policies, objectives and requirements. Authentication should be verified for Grid resource owners as well as resource requesters before they are allowed to join in scheduling activities. In order to achieve secure resource and job scheduling including minimum processing time and maximum resource utilization, A Secure Resource by using RSA algorithm on Networking and Job Scheduling model with Job Grouping strategy(JGS) in Grid Computing has been proposed. The result shows significant improvement in the processing time of jobs and resource utilization as compared to dynamic job grouping (DJG) based scheduling on smart grids (SG).

*Keywords*— RSA, SMART GRID, DYNAMIC JOB GROUPING, JOB GROUPING STRATEGY.

## 1. INTRODUCTION:

Wellner define grid technology as "the technology that enables resource virtualization, on-demand provisioning, and service (resource) sharing between organizations." IBM defines grid computing as "the ability, using a set of open standards and protocols, to gain access to applications and data, processing power, storage capacity and a vast array of other computing resources over the Internet. A grid is a type of parallel and distributed system that enables the sharing, selection, and aggregation of resources distributed across 'multiple' administrative domains based on their (resources) availability, capacity, performance, cost and users' quality-of-service requirements In 1965 Fernando Corbató and the other designers of the Multics operating system envisioned a computer facility operating "like a power company or water company". Buyya/Venugopal define grid as "a type of parallel and distributed system that enables the sharing, selection, and aggregation of geographically distributed autonomous resources dynamically at runtime depending on their availability, capability, performance, cost, and users' quality-of-service requirements". CERN, one of the largest users of grid technology, "a service for sharing computer power and data storage capacity over the Internet." Grid computing is designed to work independent problems in parallel, thereby leveraging the computer processing power of a distributed model. Prior to grid computing, any advanced algorithmic process was only available with super computers. Grid computing is a term referring to the combination of computer resources from multiple administrative domains to reach a common goal.

Ian Foster lists these primary attributes:
(i)Computing resources are not administered centrally.
(ii)Open standards are used.

GRID ARCHITECTURE:

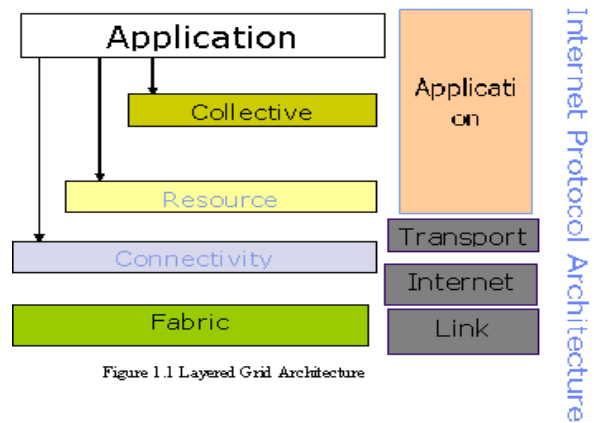

Figure 1.1 Layered Grid Architecture

In figure1.1 Grid architecture defines a layered collection of protocols.
1. Fabric layer
2. Connectivity layer
3. Resource layer
4. Collective layer

(1) Fabric layer includes the protocols and interfaces that provide access to the resources that are being shared, including computers, storage systems, datasets, programs, and networks. This layer is a logical view rather then a physical view. For example, the view of a cluster with a local resource manager is defined by the local resource manger, and not the cluster hardware. Likewise, the fabric provided by a storage system is defined by the file system that is available on that system, not the raw disk or tapes.

(2)The connectivity layer defines core protocols required for Grid-specific network transactions. This layer includes the IP protocol stack (system level application protocols transport and internet layers), as well as core Grid security protocols for authentication and authorization.

(3)Resource layer defines protocols to initiate and control sharing of (local) resources. Services defined at this level are gatekeeper, GRIS, along with some user oriented application protocols from the Internet protocol suite, such as file-transfer.

(4)Collective layer defines protocols that provide system oriented capabilities that are expected to be wide scale in deployment and generic in function. This includes bandwidth brokers, resource brokers.

**Resource management in grid computing** For running applications, resource management and job scheduling are the most crucial problems in grid computing systems. In future the development of grid technology, it is very likely that corporations, universities and public institutions will exploit grids to enhance their computing infrastructure. The goal of scheduling is to achieve highest possible system throughput and to match the application need with the available computing resources.

In figure 1.2 basic grid model generally composed of a number of hosts, each composed of several computational resources, which may be homogeneous or heterogeneous.

The four basic building blocks of grid model are user, resource broker, grid information service (GIS) and lastly resources. When user requires high speed execution, the job is submitted to the broker in grid. Broker splits the job into various tasks and distributes to several resources according to user's requirements and availability of resources. GIS keeps the status information of all resources which helps the broker for scheduling.

### GRIDSIM:

GridSim is a Java-based toolkit for modeling, and simulation of distributed resource management and scheduling for conventional Grid environment. GridSim is based on SimJava, a general purpose discrete-event simulation package implemented in Java. All components in GridSim communicate with each other through message passing operations defined by SimJava.

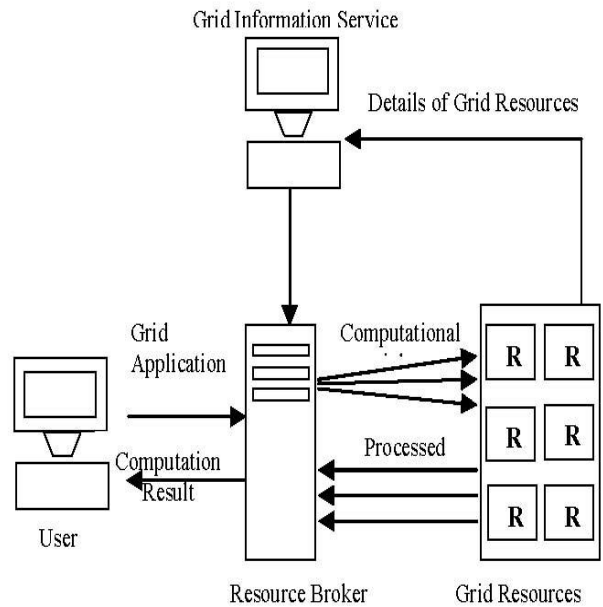

Figure 1.2 Basic grid model

### FEATURES OF GRIDSIM:

(1) It allows modeling of heterogeneous types of resources
(2) Resources can be modeled operating under space or time shared mode
(3) Resource capability can be defined in the form of MIPS(Million Instructions Per Second).
(4) Resources can be located in any time zone.
(5) Resources can be booked for advance reservation.
(6) Applications with different parallel application models can be simulated.

GridSim toolkit is suitable for application scheduling simulations in Grid Computing environment. GridSim is of great value to both students and experienced researchers who want to study Grids, or test new algorithms and strategies in a controlled environment. By using GridSim, they are able to perform repeatable experiments and studies that are not possible in a real dynamic Grid environment.

### ADVANTAGES OF USING GRIDSIM ARE

(1)Various allocation or scheduling policies can be made and integrated into GridSim easily, by extending them from one of the classes.

(2)Has the infrastructure or framework to support advance reservation, auction and Grid functionalities of a Grid system.

(3)Has the ability that reads workload traces taken from supercomputers for simulating a realistic Grid environment? This functionality is useful for testing a resource scheduling problem. This is useful for simulating over a public network where the network is congested.

Research students in the GRIDS Laboratory are themselves heavy users of GridSim and extend it whenever necessary for their own research needs. In the last 5 years, GridSim has been continuously extended in this manner to include many new capabilities and has also received contributions from external collaborators.

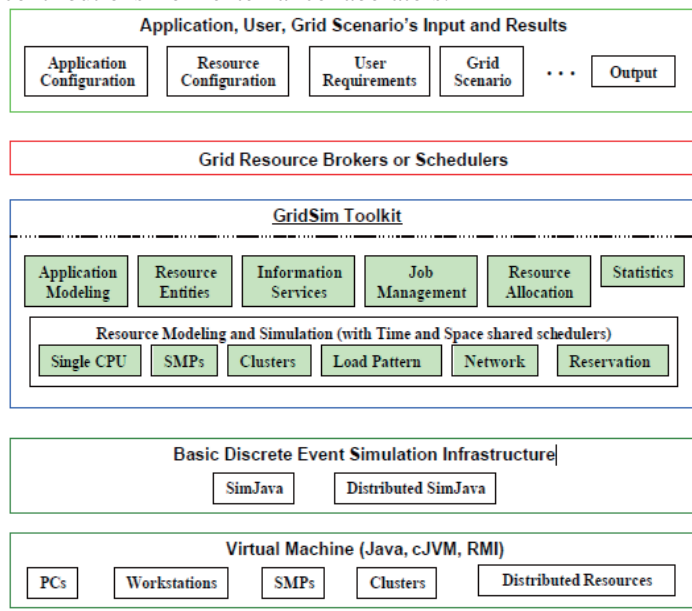

Figure 1.3. A Modular Architecture for GridSim Platform and Components

**PROBLEM FORMULATION:**

A Dynamic Job Grouping-Based Scheduling algorithm groups the jobs according to MIPS of the resource. User jobs are submitted to the scheduler and the scheduler collects the required characteristics of the available resources. It selects resources in first come first serve order. Next, it selects a specific resource and multiplies the resource processing capability specified in Million Instructions per Second (MIPS) with the granularity size, the value of this calculation produces the total Million Instructions (MI) for that particular resource to process within a particular granularity size

$$Job(MI) = Resource(MIPS) * granularity\ size$$

It selects jobs and assign to job group in first come first serve order. If the job group MI is less than that to resource MIPS than assign another job. This process continues until the resource MIPS is less to group job.

If the resource MIPS is less to group job then remove the last job MI and stop the grouping procedure. When all jobs are grouped it sends to their corresponding resource for computation.

The main disadvantage or problem of this algorithm is
(i)There is no security aspect for scheduling
(ii) Processing time is high

These disadvantages can be overcome through A secure resource and job scheduling model with job grouping strategy.

**OBJECTIVE:**

In the scheduling process, the resources are registered to the Grid Information Service (GIS). When the user submits jobs, job scheduler makes a query to GIS to obtain the information about the resource availability and its characteristics of the available resources.

The Job scheduler performs the job grouping and resource selection process. Once job groups are created, they are immediately sent for computation to the selected clusters accordingly and lastly, results are sent back to the respective Grid users

**ISSUES:**

The issue of the secure resource and job scheduling model with job grouping strategy is to overcome the deficiencies of dynamic job grouping strategy. The deficiencies are

(a)There is no security in dynamic job grouping based model

(b)The processing time to compute jobs is high in dynamic grouping based model.

In Secure resource and job scheduling model with job grouping strategy

(a)The registration process is proposed and authentication is verified for the resource requesters to join in scheduling activity

(b)Secure resource and job scheduling model with job grouping strategy contains all the Information about the resources of clusters such as MIPS (Million instructions per Second), bandwidth (in MBPS) and file size (Mb) in the global scheduler and assigns the resources to the jobs by considering all the constraints of resources. Hence the processing time of jobs will be decreased.

## 2. JOB SCHEDULING STRATEGIES

The Chapter Job scheduling strategies deals with the Researchers view of grid computing, scheduling objectives and hierarchal Job Scheduling for Cluster of Workstations

### 2.1 RESEARCHERS VIEW OF GRID COMPUTING

Grid computing is a term referring to the combination of computer resources from multiple administrative domains to reach a common goal. Foster, 2003 stated that the concept of forming pool of resources has led the way into creating large computing facilities which are currently popular known as Grid computing.

As soon as computers are interconnected and communicating, due to the faster and more capable hardware and network components and increasingly sophisticated software led to effective and efficient utilization of widely distributed resources to fulfill a range of application needs

Many researchers are working in the field of wide-area distributed computing. These research groups have implemented middleware; libraries and tools that allow the cooperative use of geographically distributed resources to act as a single powerful platform for the execution of a range of parallel and distributed applications.

This approach to computing has been known by several names, such as metacomputing, scalable computing, global computing, Internet computing and lately known as grid computing.

Grids are often constructed with the aid of general-purpose grid software libraries known as middleware. One of the main strategies of grid computing is to use middleware to divide and apportion pieces of a program among several computers, sometimes up to many thousands. Middleware is generally considered to be the layer of software sandwiched between the operating system and applications, providing a variety of services required by an application to function correctly.

According to F.Berman middleware has re-emerged as a means of integrating software applications running in distributed heterogeneous environments. In a Grid, the middleware is used to hide the heterogeneous nature and provide users and applications with a homogeneous and seamless environment by providing a set of standardized interfaces

## 2.2 SCHEDULING OBJECTIVES

Job scheduling means determining when and where the jobs are executed and how many resources are allocated. W.E.Welli stated that the basic objectives for job schedulers are to minimize response time and maximize utilization although typically there is some trade-off between both.

The goal of scheduling is to achieve highest possible system throughput and to match the application need with the available computing resources. Scheduling jobs in a Grid Computing environment is a critical problem, as it involves the usage of heterogeneous resources that are geographically distributed. Job scheduling is the process of mapping jobs to the specific available physical resources and tries to minimize the processing time of the jobs. Hence there is a need for scheduling algorithms for efficient scheduling.

According to 2010 2nd International Conference on Signal Processing Systems the scheduling algorithms are defined as
**A**. Efficient Utilization of Computing Resources Using Highest Response Next Scheduling in Grid (HRN)
**B**. Node Allocation in Grid Computing Using Optimal Resource Constraint (ORC) Scheduling
**C**. Hierarchical Job Scheduling for Clusters of Workstations (HJS)
**D**. Resource Co Allocation for Scheduling Tasks with Dependencies in grid (RCSTD)
**E** Job Schedule Model Based on grid environment(JSMB)
**F**. Dynamic Job Grouping-Based Scheduling for Deploying Applications with
Fine Grained Tasks on Global Grids (DJGBS)
**G**. Scheduling Framework for Bandwidth-Aware Job Grouping-Based scheduling in grid computing (SFBAJG)

**H**. A Bandwidth-Aware Job Grouping-Based Scheduling on Grid Environment (BAJGBS)

JOB GROUPING STRATEGY

The World Academy of Science, Engineering and Technology 64 2010 describes this grouping strategy maximizes the utilization of Grid resources, reduces processing time of jobs and network delay to schedule and execute jobs on the Grid. Grouping strategy is based on processing capability (in MIPS), bandwidth (in Mb/s), and memory-size (in Mb) of the available resources.

Jobs are grouped according to the capability of the selected resource. Therefore, the following conditions must be satisfied.
(i) Groupedjob_MI ≤ Resource_MIPS * Granularity size
(ii) Groupedjob_MS ≤ Resource_MS
(iii) Groupedjob_MS ≤ Resource_baud_rate * Tcomm

Where, MI (Million Instruction) is job's required computational power, MIPS (Million Instruction Per Second) is processing capability of the resource. Granularity size is user defined time which is used to measure total number of jobs that can be completed within a specified time, Grouped job_MS is required Memory Size of group job, Resource_MS is the amount of Memory available at resource Baud rate is the bandwidth capacity of resource, and Tcomm is the job's communication time Eq (i) required computational power of grouped job shouldn't exceed to the resource's processing capability. Eq (ii) Memory-size requirement of grouped job shouldn't exceed to the resource's memory-size capability. Eq (iii) Memory-size of the grouped job shouldn't exceed to resource's transfer capability within a specific time period.

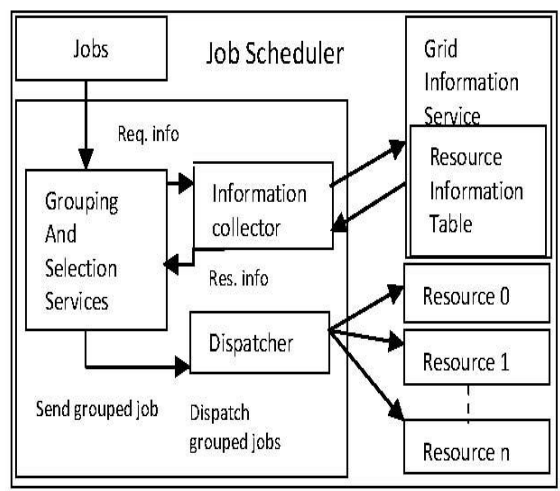

Figure2.2 Job grouping strategy

In the Figure2.2 jobs are send to the grouping and selecting services through job scheduler. Grouping and selecting services can be done by acquiring the information from the information collector. Information collector gets the information from the grid information service and the resource information table and sends the acquired information to the grouping and selection services.

After getting information the grouping and selecting services sends the grouped jobs to the dispatcher. Dispatcher dispatches the grouped jobs and finds the resources and the jobs are computed.

**Hierarchical Job Scheduling for Clusters of Workstations**

According to the Proceedings of the sixth annual conference of the Advanced School for Computing and Imaging, stated that advent of large computing power in workstations and high speed networks, the high performance computing community is moving from the use of massively parallel processors (MPPs) to cost effective clusters of workstations.Integrating a set of clusters of workstations into one large computing environment can improve the availability of computing power, e.g. Globus. The scheduling system for large environments can be divided into two levels scheduling across clusters (wide area scheduling) and scheduling within a cluster. Zhou et al introduced two level time-sharing scheduling for parallel and sequential jobs to attain a good performance for parallel jobs and a short turn-around time for sequential jobs.

At the upper level, time slots are used. Each time slot is divided into 2 slots (one for sequential jobs and a second for tasks of parallel jobs). However at the local level a task of a parallel job may share its time slot with sequential jobs. K.Y Wang uses hierarchical decision scheduling that is implemented using global and local schedulers. The global scheduler is responsible for long term allocation of system resources, and the local scheduler is responsible for short term decisions concerning processor allocation

**Implementation of Secure Resource RSA Algorithm and Job Scheduling Algorithm on Smart Grid**

Implementation deals with Secure Resource and Job scheduling Architecture, modules and System design, User and Resource registration Algorithm, Job Grouping Algorithm, Ellipse and SQLserver2008 installations. The Secure Resource and Job scheduling model is divided into three levels

    (1) user level
    (2) global level and
    (3) cluster level.

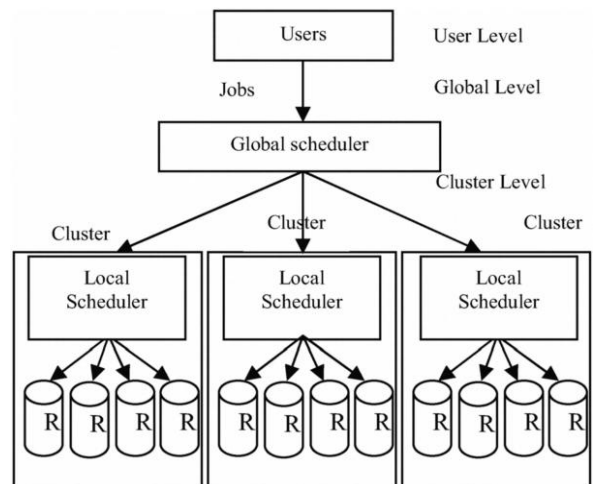

Fig Secure resource and Job scheduling model

Global level and local level scheduling are for cluster of workstation. There are two levels time-sharing scheduling for parallel and sequential jobs. Time slots are used. Each time slot is divided into 2 slots, one for sequential jobs and another for parallel jobs.

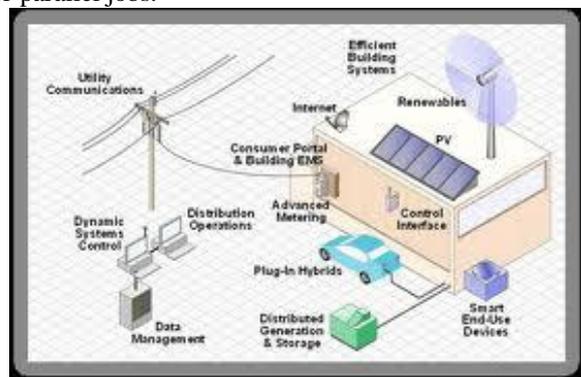

At the local level, a task of a parallel job may share its time slot with sequential jobs. Global level scheduler receives jobs and makes a queue and sends these jobs for execution according to the capability of available resources at the cluster. In cluster level, jobs are computed and results are sent back to the users.

The global and local scheduler interacts with each other to make an optimal scheduling of jobs. The local scheduler will improve the scheduling decisions made by global scheduler according to the resource availability in the cluster.

**User, Resource Registration algorithm**

Grid resources are accessible only to the authorized Grid users. Authentication should be verified for Grid resource owners as well as resource requesters before they are allowed to join in scheduling activities. Unauthorized users are not allowed to access the Grid.

In proposed model, registration is verified using user ID. If any new user arrives, registration should be completed first and then new grid user ID is assigned by the GIS.The user and registration algorithm searches for the authentication of the user. It is also used to request the resources to the scheduler. If the resource is available then the scheduler gets the resource from the Grid information system (GIS) and sends it back to the user.

**Sequence diagram for secure resource and job scheduling model with job grouping**

Sequence diagrams are used to represent or model the flow of messages, events and actions between the objects or components of a system. Sequence Diagrams are used primarily to design, document and validate the architecture, interfaces and logic of the system by describing the sequence of actions that need to be performed to complete a task. In the figure3.3 the resources needs to be registered to Grid information service. Grid user submits jobs to Admin. Admin checks for the availability of resources and requests the available resource information from the Grid information system. The resource list is provided to Admin through Grid information service. The Admin request the resource characteristics from Grid information system and gets the resource characteristics list. Admin submits the grouped jobs and result is send to the user.

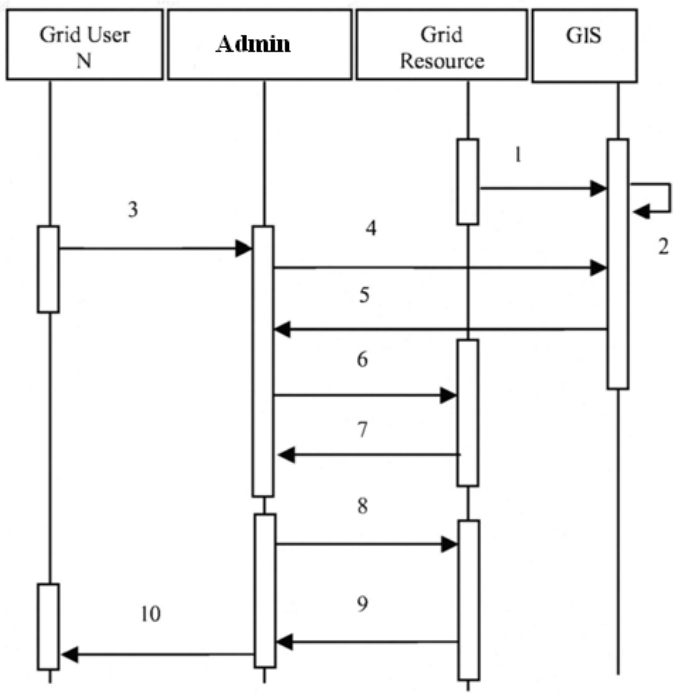

I. Register Resource to Grid Information Service (GIS)
2. Update Table of GIS
3. Submit Jobs
4. Request available Resource Information
5. Get Resource List
6. Request Resource Characteristics
7. Get Resource Characteristics
8. Submit Grouped Job I to N
9. Completed Grouped Job I to N
10. Result to user

**Sequence diagram for dynamic job grouping based scheduling**

In the figure3.4 the resources needs to be registered to Grid information service. Grid user submits jobs to Admin. Admin checks for the availability of resources and requests the available resource information from the Grid information system.

The resource list is provided to Admin through Grid information service. The Admin request the resource characteristics from Grid information system and gets the resource characteristics list.

Admin submits the grouped jobs,if grouped job(MI) is grater than resource(MI) send that job to joblist1.The job in joblist1 should be grouped and the Joblist1 needs to wait until the jobs in the joblist assigned with resources

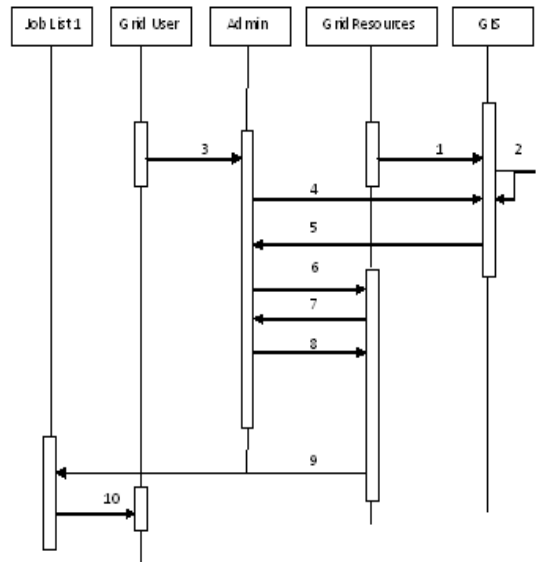

Fig sequence diagram for dynamic job grouping based scheduling

1. Register Resource to Grid Information Service (GIS)
2. Update Table of GIS
3. Submit Jobs
4. Request available Resource Information
5. Get Resource List
6. Request Resource Characteristics
7. Get Resource Characteristics
8. Submit Grouped Job I to N
9. If grouped Jobs (MI) > resource (MI) then send to Job List1
10. Then Job List1 send to Job List (Grid User)

**User resource registration algorithm**
  **Input:** Username and register resource
  **Output:** authenticates the user and checks the availability of the resources
   Step1 : User name and user id is authenticated
    var : user, resource
    for (i = 1; i<Nu; i++)
    {Nu = Number of users.//}
    getrequest(specification);

```
        //specification is a tuple
        <user name, user id, user time .... .//
        Authenticated();
        getresponse(user_ID)
        }
```

Step2: Resource name, Resource enter time, Resource MIPS, Resource bandwidth,
      Resource memory size are registered to database
```
      for (i = 1; i<Nr; i++)
      { // Nr = Number of resource//}
       request register to GIS;
       getrequest(specification);
       specification is a tuple
       <resource name, resource enter time, resource MIPS, resource bandwidth, resource
       Memory size>
       Authenticated();
       }
```
Step3: Scheduler checks the availability of resources
```
      Submit job to scheduler ()
      { scheduler get information of resource from GIS();
      while resource = 'available'
```
Step4: If resources are available executes the job grouping algorithm
```
      call job grouping algo();
       scheduler submits job to resource;
       resource computed jobs;
       end back result to users;
       }
      }
```

Job grouping algorithm
  Input   : Joblength
  Output  : Resources and clusters
  Step1 : Submits the user jobs to the grid
  Step2 : Receive job list, JList
  Step3 : Receive resource list, Rlist
      for i:= 0 to JobList Size-1 do
      for j:=O to Rlist_Size-1 do
  Step4 : Grouped Job(million instructions) should not be equal to zero
      GroupedjobMI:= 0
  Step5: Resource Million instructions can be obtained by multiplying resource MIPS
          (processing capability) with granularity size which is taken as 3 in the project
      R(mi)J:= Rlist_MIPs* Granularity Size
  Step6: If Grouped job million instructions is less to Resource million instructions
          Add another job
      while GroupedjobMI < R(mi)J and i < JobList_Size-1 do
      Groupedjob(MI):=Groupedjob(MI)+joblisti_(MI)
       i++
  Step7: If Grouped job million instructions is greater than Resource million instructions
      if GroupedjobMI > R(mi)J then

  Step8: The loop is repeated between the available resources and should assign the resource.
  Step9: Place the Grouped job(j )to Target Resource List(j )for computation
  Step10: Receive computed Grouped job from Resource List j.

RSA Algorithm for Smart Grids Security:
```
RSAParameters m_private;
RSAParameters m_public;
public byte[] GetHash(byte[] message)
    {
        MD5CryptoServiceProvider md5 = new MD5CryptoServiceProvider();
        return md5.ComputeHash(message, 0, message.Length);
    }
    public byte[] CreateSignature(byte[] hash)
    {
    RSACryptoServiceProvider RSA = new RSACryptoServiceProvider();
    RSAPKCS1SignatureFormatterRSAFormatter=new; RSAPKCS1SignatureFormatter(RSA);
    RSAFormatter.SetHashAlgorithm("MD5");
    m_public = RSA.ExportParameters(false);
    m_private = RSA.ExportParameters(true);
    return RSAFormatter.CreateSignature(hash);
    }
    public bool VerifySignature(byte[] hash, byte[] signedhash)
    {
    RSACryptoServiceProvider RSA = new RSACryptoServiceProvider();
    RSAParameters RSAKeyInfo = new RSAParameters();
    RSAKeyInfo.Modulus = m_public.Modulus;
    RSAKeyInfo.Exponent = m_public.Exponent;
    RSA.ImportParameters(RSAKeyInfo);
    RSAPKCS1SignatureDeformatter RSADeformatter = new RSAPKCS1SignatureDeformatter(RSA);
    RSADeformatter.SetHashAlgorithm("MD5");
    return RSADeformatter.VerifySignature(hash, signedhash);
    }
```

**Experimental results**

In the Chapter "Secure resource and a job scheduling model with a job grouping strategy  has to provide security for the grid information system.

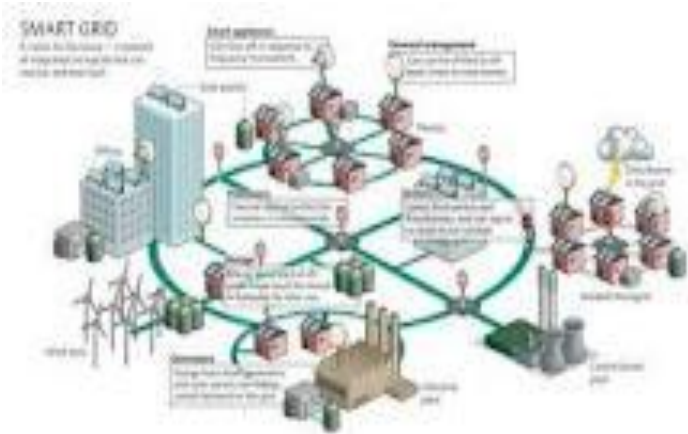

- Admin or the server has to validate the user id and password of the user.
- A new user must register before it should enter into scheduling process.
- After the registration process the user can submit jobs, view profile and view assigned resources.
- The admin can view profiles, view user jobs, View resources, Assign resources, view assigned resources, assign cluster, View cluster and finds the processing time of secure resource and job scheduling algorithm and dynamic job grouping algorithm.
- The profile that is entered by a new user will be updated in the Admin module.
- Resources MIPS, resource bandwidth, name of the resource and resource enter time and resource memory size are registered to database.
- The admin will also provide the cluster to process the group of jobs.
- To process these jobs the Secure resource and a job scheduling model (SRJM) takes less time comparing to dynamic job grouping based algorithm.(DJGB).
- In the module of SRJM, contains the block of SRJM and DJGB.
- SRJM is secured and it can be accessed by the authorized user.
- The processing time of jobs in SRJM is less compared to DJGB.
- The SRJM algorithm is written in java in the project.
- A simulation is conducted in heterogeneous environment where each cluster has machines with different characteristics and MIPS, to verify the improvement of proposed model over other dynamic job grouping based scheduling algorithm.
- In this simulation, size of the granularity is taken as 3 seconds for both scheduling algorithm.

| RESOURCE NAME | RESOURCE MIPS | RESOURCE BANDWIDTH | RESOURCE MEMORY SIZE | GRANULARITY SIZE |
|---|---|---|---|---|
| R1 | 10 | 100 | 100 | 3 |
| R2 | 20 | 150 | 200 | 3 |
| R3 | 30 | 200 | 300 | 3 |
| R4 | 40 | 250 | 400 | 3 |

TABLE REGISTERED RESOURCES (R1 TO R4)

| RESOURCE NAME | RESOURCE MIPS | RESOURCE BANDWIDTH | RESOURCE MEMORY SIZE | GRANULARITY SIZE |
|---|---|---|---|---|
| R5 | 50 | 300 | 500 | 3 |
| R6 | 60 | 350 | 600 | 3 |
| R7 | 70 | 400 | 700 | 3 |
| R8 | 80 | 450 | 800 | 3 |

TABLE REGISTERED RESOURCES (R4 TO R8)

| RESOURCE NAME | RESOURCE MIPS | RESOURCE BANDWIDTH | RESOURCE MEMORY SIZE | GRANULARITY SIZE |
|---|---|---|---|---|
| R9 | 90 | 500 | 900 | 3 |
| R10 | 100 | 550 | 1000 | 3 |
| R11 | 110 | 600 | 1100 | 3 |
| R12 | 120 | 650 | 1200 | 3 |

TABLE REGISTERED RESOURCES (R8TO R12)

| RESOURCE NAME | RESOURCE MIPS | RESOURCE BANDWIDTH | RESOURCE MEMORY SIZE | GRANULARITY SIZE |
|---|---|---|---|---|
| R13 | 130 | 700 | 1300 | 3 |
| R14 | 140 | 750 | 1400 | 3 |
| R15 | 150 | 800 | 1500 | 3 |
| R16 | 160 | 850 | 1600 | 3 |

Table registered resources (R12to R16)

**Comparison between the algorithms according to their processing time**

The proposed secure resource and Job grouping based scheduling provides significant improvement in decreasing the processing time of jobs compared to dynamic job grouping strategy

| NO of Jobs | processing time | |
|---|---|---|
| | SRJM | DJGB |
| 3 | 4 | 10 |
| 5 | 7 | 22 |
| 8 | 11 | 28 |
| 10 | 16 | 34 |
| 14 | 21 | 40 |

**Table Comparison between the algorithms**

**Conclusion and future scope:**

In order to achieve secure resource and job scheduling including minimum processing time and maximum resource utilization, A Secure Resource RSA algorithm and Job Scheduling model with Job Grouping strategy in Grid Computing has proposed.

The Secure Resource and job scheduling algorithm will show better comparative result than dynamic job grouping based scheduling algorithm. The simulation environment has shown the proposed model is able to perform job and resource scheduling in grid environment and provide real grid Computing environment.

In future this work can be extended to implement job resource scheduling model with load balancing to increase the performance of Smart grid system.

## References

[1] F. Berman, G. Fox, and T. Hey," "Grid computing: Making the Global Infrastructure a Reality", Wiley and sons, 2003.
[2] Berman, F. , Fox, G. and Hey, A. " Grid Computing – Making the Global Infrastructure a Reality. London, Wiley, 2003
[3] R. Buyya and M. Murshed, GridSim; A toolkit for the modeling and simulation of distributed management and scheduling for grid computing, 2002
[4] D. L. Clark, J. Casas, W. Otto, M. Prouty, and J. Walpole, "Scheduling of parallel jobs of parallel jobs on dynamic", heterogeneous networks,1995.
[5] Foster, C. Kessalman and S. Tuecke,, "The taxonomy Enabling scalable virtual organization", lnt. J.Supercomputing, vol. 15, no. 3, pp. 200-222, 2001.
[6] A. Hori, H. Tezuka, Y. Ishikawa, N. Soda, H. Konaka, and M. Maeda "Implementation of gang scheduling on workstation cluster", Springer, In Job Scheduling Strategies for Parallel Processing, vol. 1996 of LNCS, Berlin, 1996.
[7] Ms.P.Muthuchelvi, Dr.V.Ramachandran, "ABRMAS: Agent Based Resource Management with Alternate Solution," IEEE, The Sixth International Conference on Grid and Cooperative Computing, GCC 2007.
[8] Nithiapidary Muthuvelu, Junyang Liu, "A Dynamic Job Grouping-Based Scheduling for Deploying Application with Fine-Grained tasks on Global Grids", vol. 44, Australasian Workshop on Grid Computing and e-Research, AusGrid -2005.
[9] J. Santoso, G.D. van Albada, "Hierarchical Job Scheduling for Clusters of Workstations", In ASCI, proceedings of the Sixth annual conference of the Advanced School for Computing and Imaging, pp. 99-105, 2000
[10] K. Y. Wang, D. C. Marinescu, and o. F. Carbunar, " Dynamic scheduling of process groups", Concurrency: Practice and Experience, I OC 4):265-283, April 1998.
[11] B. Zhou, R. Brent, D.Walsh, and K. Suzaki, " Job scheduling strategies for networks of workstations", In Job Scheduling Strategies for Parallel Processing, vol. 1459 of LNCS, pp. 143- 157, Berlin, 1998. Springer.

**Bibliography**

**Mr. P. Radha Krishna Reddy** received his B.Sc (CS) from Sri Venkateswara University-Tirupati. M.Sc in Computer science from Sri Venkateswara University-Tirupati, and pursuing M.Tech in Computer Science and Engineering from Vaagdevi Institute of Technology and Sciences, JNTU-Anantapur. He has 2+ years of teaching Experience. He has the membership in INTERNATIONAL ASSOCIATION OF ENGINEERS and COMPUTER SCIENCE JOURNALS.

**Mr. Ashim Roy** received his B.Tech in Information Technology from Siliguri Institute Of Technology in the year 2006,affiliated to West Bengal University Of Technology . Pursuing M.Tech in Information Technology from I Tech College Falakata(West Bengal) affiliated to University Of Karnataka.

**Ms. G.Sireesha** received her B.Tech in Computer science and Engineering from Royal Institute of Technology and Science, JNTU, Hyderabad, M.Tech in Computer science (Parallel computing) from Aurora's Engineering College, JNTU, Hyderabad.She is working as a Assistant Professor in Computer Science and Engineering in Guru Nanak Institute of Engineering & Technology, JNTU-Hyderabad.

**Ms. Ismatha Begum** received her B.Tech from Sana engineering college from JNTUH. M.Tech in parallel computing from aurora's engineering college. worked as assist prof in sri devi womens engineering college from 2008 -2009 and worked as asst prof in vastalya engineering college from 2010 to 2011.

**Mr. S.Siva Ramaiah** received his MCA from Sri Venkateswara University-Tirupati. M.Tech in Computer Science and Engineering from Acharya Nagarjuna University. He has 7 years of teaching Experience. He is the HOD of MCA & MBA in Vaagdevi Institute of Technology and Sciences, JNTU-Anantapur